\documentclass[manuscript,screen]{acmart}

\AtBeginDocument{%
  \providecommand\BibTeX{{%
    \normalfont B\kern-0.5em{\scshape i\kern-0.25em b}\kern-0.8em\TeX}}}

\begin{document}

\title{Efficient Quantum Circuit Simulation by Tensor Network Methods on Modern GPUs}

\author{Feng Pan}
\authornote{Both authors contributed equally to this research.}
\affiliation{%
  \institution{CAS Key Laboratory for Theoretical Physics, Institute of Theoretical Physics, Chinese Academy of Sciences}
  \city{Beijing}
  \country{China}
  \postcode{100190}
}
\affiliation{
    \institution{Centre for Quantum Technologies, National University of Singapore}
    \postcode{117543}
    \country{Singapore}
}
\email{fengpan1994@gmail.com}
\author{Hanfeng Gu}
\authornotemark[1]
\affiliation{%
  \institution{NVIDIA}
  \city{Shanghai}
  \country{China}
}
\email{henrygu@nvidia.com}

\author{Lvlin Kuang}
\affiliation{%
  \institution{NVIDIA}
  \city{Beijing}
  \country{China}
}
\email{lkuang@nvidia.com}
\author{Bing Liu}
\affiliation{%
  \institution{NVIDIA}
  \city{Shenzhen}
  \country{China}
}
\email{petrickl@nvidia.com}

\author{Pan Zhang}
\affiliation{%
  \institution{CAS Key Laboratory for Theoretical Physics, Institute of Theoretical Physics, Chinese Academy of Sciences}
  % \streetaddress{P.O. Box 1212}
  \city{Beijing}
  \country{China}
  \postcode{100190}
}
\affiliation{%
  \institution{School of Fundamental Physics and Mathematical Sciences,
 Hangzhou Institute for Advanced Study, UCAS}
  % \streetaddress{P.O. Box 1212}
  \city{Hangzhou}
  \state{Zhejiang}
  \country{China}
  \postcode{310024}
}
\affiliation{%
  \institution{International Center for Theoretical Physics Asia-Pacific}
  % \streetaddress{P.O. Box 1212}
  \city{Beijing/Hangzhou}
  \country{China}
}
\email{panzhang@itp.ac.cn}

\renewcommand{\shortauthors}{Pan and Gu, et al.}

\begin{abstract}
  Efficient simulation of quantum circuits has become indispensable with the rapid development of quantum hardware. The primary simulation methods are based on state vectors and tensor networks. As the number of qubits and quantum gates grows larger in current quantum devices, traditional state-vector based quantum circuit simulation methods prove inadequate due to the overwhelming size of the Hilbert space and extensive entanglement. Consequently, brutal force tensor network simulation algorithms become the only viable solution in such scenarios. The two main challenges faced in tensor network simulation algorithms are optimal contraction path finding and efficient execution on modern computing devices, with the latter determines the actual efficiency. In this study, we investigate the optimization of such tensor network simulations on modern GPUs and propose general optimization strategies from two aspects: computational efficiency and accuracy. Firstly, we propose to transform critical Einstein summation operations into GEMM operations, leveraging the specific features of tensor network simulations to amplify the efficiency of GPUs. Secondly, by analyzing the data characteristics of quantum circuits, we employ extended precision to ensure the accuracy of simulation results and mixed precision to fully exploit the potential of GPUs, resulting in faster and more precise simulations. Our numerical experiments demonstrate that our approach can achieve a 3.96x reduction in verification time for random quantum circuit samples in the 18-cycle case of Sycamore, with sustained performance exceeding 21 TFLOPS on one A100. This method can be easily extended to the 20-cycle case, maintaining the same performance, accelerating by 12.5x compared to the state-of-the-art CPU-based results and 4.48-6.78x compared to the state-of-the-art GPU-based results reported in the literature. Furthermore, the strategies presented in this article hold general applicability for accelerating problems that can be tackled by contracting tensor networks such as graphical models and combinatorial optimizations.
\end{abstract}

%    <concept>
%        <concept_id>10010147.10010341.10010370</concept_id>
%        <concept_desc>Computing methodologies~Simulation evaluation</concept_desc>
%        <concept_significance>500</concept_significance>
%        </concept>
%    <concept>
%        <concept_id>10010405.10010432.10010441</concept_id>
%        <concept_desc>Applied computing~Physics</concept_desc>
%        <concept_significance>500</concept_significance>
%        </concept>
%    <concept>
%        <concept_id>10010583.10010717.10010721.10010725</concept_id>
%        <concept_desc>Hardware~Simulation and emulation</concept_desc>
%        <concept_significance>300</concept_significance>
%        </concept>
%    <concept>
%        <concept_id>10010147.10010169.10010170.10010174</concept_id>
%        <concept_desc>Computing methodologies~Massively parallel algorithms</concept_desc>
%        <concept_significance>300</concept_significance>
%        </concept>
%  </ccs2012>

% \ccsdesc[500]{Computing methodologies~Simulation evaluation}
% \ccsdesc[500]{Applied computing~Physics}
% \ccsdesc[300]{Hardware~Simulation and emulation}
% \ccsdesc[300]{Computing methodologies~Massively parallel algorithms}



\maketitle

\section{Introduction}
\label{section:intro}

Quantum computing is a rapidly developing interdisciplinary field that combines concepts from physics and computer science. Unlike classical computers, which rely on binary bits, quantum computers use quantum mechanical systems constructed by qubits that hold the power of superpositions and entanglement. This potential of quantum computing allows to solve certain problems polynomially or exponentially faster than classical computing, making it a competitive challenger to the extended Church-Turing thesis and a promising candidate for post-Moore computation.

% possible quantum algorithms that can demonstrate advantage but currently not available
The concept of quantum computing was first introduced by pioneers such as Feynman in the 1980s~\cite{feynman1982simulating}. Since then, quantum computers have been built in various platforms. 
However, in the current Noisy Intermediate-Scale Quantum (NISQ) era~\cite{preskill2018quantum}, several tasks with rigorous theoretical foundations, such as factoring~\cite{shor1994algorithms} and database searching~\cite{grover1996fast}, cannot be executed on current quantum devices due to the inevitable noise.
% Although some experiments have proposed the construction of noiseless logical qubits using current noisy quantum devices~\cite{google2023suppressing, zhao2022realization}, practical quantum devices available today have only a few dozen to a few hundred qubits, which exhibit noticeable noise that cannot be neglected.

% current avaible quantum computational advantage experiments
To identify possible gaps between quantum and classical computing on noisy quantum devices, researchers have focused on sampling-based quantum computing tasks, such as Gaussian Boson sampling (GBS)~\cite{aaronson2011computational, hamilton2017gaussian} and Random Quantum Circuit Sampling (RCS)~\cite{aaronson2016ComplexityTheoretic, boixo2018Characterizing}. The noise-free versions of these tasks have been proven to be exponentially hard for classical computation. Several quantum hardware systems have achieved these computational tasks~\cite{arute2019Quantum, zhong2020quantum, wu2021Strong, zhu2021Quantum, madsen2022quantum}. Their results indicate quantum computational advantages over specific classical algorithms run on state-of-the-art supercomputers.

% drawbacks and limitations of current quantum supremacy experiments
The claim of quantum computational advantage based on RCS experiments on NISQ devices has been disputed since Google claimed in 2019~\cite{arute2019Quantum} that classical simulation of their 53-qubit, 20-cycle Sycamore circuits would take supercomputers 10,000 years. The Linear Cross-Entropy Benchmark (LXEB), which determines the fidelity of their sampling results, has been targeted by several spoofing algorithm~\cite{barak2020spoofing, pan2022Simulation, gao2021Limitations}, due to the extremely low LXEB value of Sycamore circuits (around $0.2\%$). Moreover, a recent seminal study has established that RCS experiments with a constant error rate per gate in the asymptotic limit are poly-time classically simulatable~\cite{aharonov2022Polynomialtime}. Although this algorithm cannot handle finite-size cases due to its exponentially large coefficient with the depolarizing error rate of each layer in the circuit, it has effectively disproven the possibility of an expanding quantum advantage in noisy quantum circuits with larger sizes.

%%  Classical simulation (simulate & validate) *** - Pan
% Why classical simulation and classical verification is important
In the meantime, classical simulation and verification of finite-size RCS experiments are also very crucial. Classical simulation involves sampling bitstrings with bounded LXEB values from quantum circuits, which helps to understand gaps between quantum and classical computing regarding this task. Classical verification, on the other hand, entails calculating the exact amplitudes sampled in the RCS experiments, serving two primary objectives. Firstly, it validates the authentic outputs of the RCS experiments, thus ensuring the reliability of the experimental results. Secondly, efficient verification methods enable the realization of numerous potential applications of such sampling-based quantum algorithms, such as solving subgraph problems~\cite{arrazola2018using} and generating random numbers with certified randomness~\cite{bassirian2021certified,aaronson2023certified}.

% can find potential applications in various other domains where the calculation of expectation values on specific bitstrings sampled in experiments is necessary~\cite{peruzzo2014variational, farhi2014quantum}. The exact computation of the sampled bitstring amplitudes can provide a more precise evaluation of these expectation values.

% traditional classical simulation methods and why they fail at simulating quantum supremacy experiments, introduce to tensor network simulation algorithms
Classical simulation and verification techniques for quantum algorithms have traditionally relied on state vector based methods. However, they are only practical for circuits with a limited number of qubits, where it is possible to store the entire state vector or density matrix. For quantum circuits with low entanglement, state vectors can be accurately approximated and efficiently operated~\cite{zhou2020What, cheng2021simulating}. These approaches result in a polynomial computational cost and enable simulation of larger system sizes. However, when attempting to perform classical simulations of quantum algorithms that demonstrate quantum computational advantages, these methods fail due to the large number of qubits (usually greater than 50) and high-entanglement circuit architecture. 

A practical way to simulate such quantum experiments is the tensor network simulation algorithm~\cite{markov2008Simulating}. This method converts all components of the quantum circuit into tensor networks, which are then contracted to obtain specific final state bitstring amplitudes. Although this approach can handle cases with a large number of qubits and high entanglement, it still faces several challenges, with two core ones being the contraction path finding and efficient implementation on current classical devices. 

Due to the substantial number of tensors involving multiple dimensions within the tensor network simulation, identifying an optimal contraction path presents an immensely challenging combinatorial optimization problem with no known efficient heuristic algorithm. Recently, various path-finding approaches employing graph-partitioning, simulated-annealing-like, and architecture-aware methods~\cite{chen2018Classical, gray2021Hyperoptimized, huang2021Efficient, pan2022Simulation, pan2022Solving} have demonstrated significant advancements in the bounded-fidelity simulation of RCS experiments~\cite{huang2021Efficient, liu2021Closing, pan2022Solving}. These advancements have significantly improved the simulation time for a 53-qubit, 20-cycle Sycamore circuit, reducing the original estimated time of 10,000 years in Google's paper to a duration comparable to quantum experiments.

Furthermore, achieving an efficient implementation of such tensor network simulation tasks poses a non-trivial challenge, as modern high-performance computing devices are not specifically designed to handle tensor contractions with the arbitrariness of dimension and contraction indices, and the computational tasks require a substantial amount of both computing capacity and memory speed. Previous studies~\cite{huang2021Efficient, pan2022Solving, liu2021Closing, kalachev2021Classical, liu2022Validating} indicate that current implementations only utilize around 20\% of the peak performance of the devices. Considering that the verification tasks are considerably more arduous than the bounded-fidelity simulation tasks, achieving a much more efficient implementation on modern devices is imperative.

The main focus of this paper is on the optimization of tensor network methods used in simulating and verifying quantum circuit-based experiments, with a view to making them more efficient on modern GPUs. Specifically, we optimize the simulation method proposed in~\cite{pan2022Solving}, and target RCS experiments. The contributions of our work can be summarized as follows: 

\begin{itemize}
    \item We devise a strategy that converts the most time-consuming component of the sparse-state tensor network simulation from sparse Einstein summation (einsum) to GEMM operations. This optimization results in a significant decrease in the overall running time of that component.

    \item We undertake an in-depth investigation of the impact of data types on the simulation results of quantum circuits. We propose a scheme that employs different precisions in different simulation stages. This approach leverages Tensor Cores on modern GPUs while ensuring accuracy and consistency of simulation results.

    \item We conduct extensive numerical experiments to test our optimization strategies. With the use of NVIDIA A100 GPUs, we successfully verify RCS experiments based on 53-qubit 18-cycle Sycamore circuits with $2^{20}$ sampled bitstrings, using 8435 GPU hours. This is 3.96 times faster than the non-optimized version. Moreover, for 53-qubit 20-cycle Sycamore circuits with three million sampled bitstrings verification task, the estimated computational resource is 1.26 million GPU hours. This is 12.5 times faster than the supercomputer's time reported in~\cite{liu2022Validating}, assuming the same single-precision peak performance is maintained. Using the proposed methodology, we could also achieve the simulation task of bounded LXEB value ($0.2\%$), at an estimated computational cost of 2819 GPU hours. Compared to the results reported in~\cite{kalachev2021Classical} and~\cite{pan2022Solving} with the same computing complexity, the simulation could be accelerated by 4.48 and 6.78 times, respectively.
\end{itemize}

%% applications

It is worth noting that the algorithm employed in this study does not make assumptions on the number of qubits, the types of two-qubit gates, or the quantum circuit connections. As a result, the optimizations described in this article are generalizable to any quantum circuits and any quantum circuit based algorithm~\cite{peruzzo2014variational, farhi2014quantum}, with the circuits used in random circuit sampling representing particularly challenging cases. Furthermore, these optimizations can be extended to address other problems that involve tensor network contractions, including graphical models~\cite{robeva2017Duality}, combinatorial optimization problems~\cite{kourtis2019Fast}, quantum error correction~\cite{bravyi2014Efficient}, and so forth.

%%  paper structure

The remainder of the paper is structured as follows: Section~\ref{section:bg} provides an overview of quantum circuit models, tensor networks, RCS experiments, and simulations on GPUs. In Section~\ref{section:algorithm}, we provide a detailed description of the algorithms we employ to leverage tensor network contractions for simulating quantum circuits. Section~\ref{section:opt on GPU} presents a comprehensive explanation of the enhancements we introduce to the tensor network simulation to significantly enhance its efficacy on modern GPUs. In Section~\ref{section:experiments}, we assess the performance of the optimized tensor network simulation and compare it to other implementations. We wrap up the study by discussing our conclusions and future research directions in Section~\ref{section:conclusions}.

\section{Background}
\label{section:bg}
%%  How quantum circuit maps to tensor network - Pan
%%  RCS experiments background (how to do, benchmarking(XEB), maps to TN contraction) - Pan
%%	• On different platform. (why GPU?  opts )  - Pan
%%	• Methods on GPU (Tensor Core)  - Henry
%%	• Efficient tensor contraction on GPU (TTGT and GEMM-like, cuTensor)  - Henry
%%

\subsection{Quantum circuits and Tensor Network Simulation}
Quantum circuits, like their classical counterparts, are circuits made up of quantum gates. In contrast to classical circuits that utilize high and low voltages to represent binary states, quantum circuits use quantum superposition to encode multiple classical states at the same time. The interaction and manipulation of quantum states are governed by unitary operations called quantum gates. To extract information encoded within quantum states, measurements are performed. The result of a measurement is a collapsed 0-1 bitstring in the measurement basis. It is noteworthy that measurements destroy the superposition of quantum states. Thus, repetition of the entire experiment is necessary to gain more information regarding the quantum state.

Quantum gates and initial quantum states can be mathematically represented by tensors and vectors, respectively. The interconnection and ordering of quantum gates and quantum states in quantum circuits can be mapped to the bonds between tensors. This mapping enables the conversion of a quantum circuit into a unique tensor network. Given a separable initial state, the tensor network input is closed in the initial state section and opened in the final state section. The open bonds found in this area indicate the configurations of individual qubits. By contracting the tensor network, we may acquire a vector that combines all open bonds into a singular entity. This vector comprises the final quantum state of the circuit and contains all amplitudes of the entire Hilbert space. An illustration of a tensor network representation of a quantum circuit is shown in Figure~\ref{fig:simulation demo}(a).

The measurement process in quantum circuits lacks a corresponding tensor network methodology for large-scale quantum circuits. This is due to the fact that the quantum state vector cannot be stored in its entirety by any classical computer when the number of qubits surpasses 50. As a result, it is impossible to generate samples from it. For tensor network simulations, even though the entire state vector cannot be retrieved, the amplitudes of small sets of bitstrings can be obtained. By utilizing these bitstrings and their corresponding probabilities, sampling approaches such as importance sampling or rejection sampling can be employed to extract samples from them. This process captures the measurement process of quantum experiments~\cite{boixo2018Characterizing, huang2021Efficient}.

\subsection{Random Quantum Circuit Sampling Experiments}
RCS experiments\cite{aaronson2016ComplexityTheoretic, boixo2018Characterizing,arute2019Quantum,wu2021Strong,zhu2021Quantum} execute multiple measurements on the final state of a random quantum circuit, generating bitstring samples that align with a specific distribution called the Port-Thomas distribution~\cite{porter1956fluctuations} of that final state. The construction of the quantum circuit is typically achieved by organizing sequences of quantum gates, with each sequence composed of a single-qubit gate layer, followed by a two-qubit gate layer. The entire circuit is comprised of numerous such cycles.

In order to make classical simulation difficult, two-qubit quantum gates are carefully designed to be full rank in the temporal direction within random quantum circuits. This design choice increases classical complexity. The quantum circuit's randomness comes from the utilization of a specific gate set for single-qubit gates, which are randomly selected. In the designs of two quantum circuits that have demonstrated quantum computational advantage \cite{arute2019Quantum, wu2021Strong, zhu2021Quantum}, fermionic simulation (fsim) gates with carefully tuned parameters are used as two-qubit gates, while single-qubit gates are selected randomly from the $\sqrt{X}, \sqrt{Y},$ and $\sqrt{W}$ gates.

Upon obtaining the sampled bitstrings, a LXEB procedure is then executed to evaluate the quality of said samples. LXEB is introduced as a viable substitute for the actual fidelity, which is practically difficult to obtain in quantum circuits that involve a large number of qubits. The LXEB equation is, 

\begin{equation}
F_l = 2^N\sum_x p_U(x)q(x) - 1 = 2^N\mathrm{E}_{x\sim q(x)}p_U(x) - 1,
\end{equation}
where $N$ is the number of qubits in the quantum circuit, $x$ denotes the bitstring, and the summation is performed over all possible bitstrings. $p_U(x)$ represents the probability of said bitstring in the ideal circuit that is devoid of noise, and $q(x)$ represents the probability of said bitstring in the real circuit.

However, in real-world quantum experiments, obtaining a complete ensemble of bitstrings for the evaluation of $q(x)$ and calculation of its associated expectation value is infeasible. Therefore, a set number of bitstrings are randomly sampled, with the LXEB subsequently computed as

\begin{equation}
\label{eq:lxeb_2}
F_l = 2^N\sum_{i=1}^m p_U(x_i)/m - 1.
\end{equation}

It is worth noting that $p_U(x_i)$ should be calculated using classical computation, but due to the difficulty of calculating it, quantum experiments will be done in simplified circuits which can be simulated efficiently in classical computers. The circuits for demonstration of quantum computational advantages only differ in their choice of two-qubit gates, allowing LXEB values to be inferred from those of simpler circuits.

Quantum hardware has demonstrated the capacity to perform RCS experiments with a high level of efficiency. A single instance of quantum state preparation and measurement can be completed in a mere few milliseconds. The Sycamore circuits, for example, required approximately 600 seconds to sample 3 million bitstrings, with a final LXEB of $0.224\%$. The largest Zuchongzhi circuit, by contrast, necessitated 4.2 hours to sample 70 million bitstrings, but exhibited a final LXEB of $0.0366\%$.

\subsection{Modern GPU} %% Henry
%% demonstrate modern GPU architecture, related hpc libraries

%% note GPU is not only for NV, should declare that

% GPUs are designed with a single instruction multiple thread (SIMT) architecture, which enables them to execute numerous parallel computing tasks concurrently. This feature renders them well-suited for handling compute-intensive tasks. Moreover, GPUs boast large memory and high memory bandwidth (often exceeding TB/s), facilitating the efficient management of memory-intensive tasks. Given these qualities, GPUs are a preferred choice for implementing tensor network simulations.

Modern GPUs are SIMT architectures that break down computing tasks into thousands of threads for parallel execution, making it highly suitable for compute-intensive tasks. The bandwidth of GPU memory can typically reach TB/s levels, giving GPU a significant advantage in processing tasks that require high bandwidth. Additionally, modern GPU typically has large memory, which is crucial for tensor network contraction.

% \begin{figure}[h]
%     \centering
%     \includegraphics[scale=0.18]{figs/chap2_ga100_sm.png}
%     \caption{The GA100 streaming multiprocessor (source: NVIDIA).}
%     \label{fig:chap2_ga100}
%     % \Description{}
% \end{figure}

In modern GPU hardware represented by NVIDIA GPUs, the core computing unit is the streaming multiprocessor (SM), which contains CUDA Cores (FP64, FP32, INT32) and Tensor Cores. These units perform tasks in three parallel computing levels, grid, block, and thread. The grid is the largest unit, representing the entire program or task; the block is a subset of the grid, representing a group of threads that can collaborate; the thread is an element within the block, representing the smallest unit that executes specific instructions.

\begin{scriptsize}
\begin{table}
\centering
  \caption{Floating-point format and peak performance on A100 GPU}
  \label{tab:chap2_a100__format_perf}
  % \begin{tabular}{p{2.5cm}p{1.2cm}p{1cm}p{1cm}p{1cm}p{1cm}}
  \begin{tabular}{lccccc}
    \hline
    Floating Precision & FP64 & FP32 & TF32 & FP16 & BF16  \\
    \hline
    Sign Bits & 1 & 1 & 1 & 1 & 1 \\
    Exponent Bits & 11 & 8 & 8 & 5 & 8 \\
    Mantissa Bits & 52 & 23 & 10 & 10 & 7 \\
    Range Min & E-308 & E-38 & E-38 & E-8 & E-38 \\
    Range Max & E+308 & E+38 & E+38 & E+4 & E+38 \\
    Significant Digits & 15$\sim$16 & 7$\sim$8 & 3$\sim$4 & 3$\sim$4 & 2$\sim$3 \\
    \hline
    CUDA Core(TFlops) & 9.7 & 19.5 & / & 78 & 39 \\
    Tensor Core(TFlops) & 19.5 & / & 156 & 312 & 312 \\
    \hline
\end{tabular}
\end{table}
\end{scriptsize}

The Tensor Core, which was first introduced in the NVIDIA Tesla V100 GPU and further enhanced in recent NVIDIA Ampere and Hopper GPUs~\cite{choquette2021nvidia, elster2022nvidia}, is a specialized unit for matrix multiplication and accumulation operations. For these operations, the Tensor Core is more efficient than the CUDA Core since the Tensor Core can perform multiple fused multiply–add (FMA) operations per clock cycle per instruction, while the CUDA Core can only perform one per clock cycle per instruction. Thus, the Tensor Core can provide up to 2--8x speedups compared to the CUDA Core (see Table~\ref{tab:chap2_a100__format_perf}). However, the Tensor Core has hardware limitations and can only process matrix operations of fixed sizes. To fully utilize the computing power of Tensor Cores, each dimension of the input matrix must be a multiple of 8~\cite{nvidia2023cublas}. This requires branch merging~\cite{huang2021Efficient} in tensor network contraction, as quantum gates (single-qubit or two-qubit) are usually represented as 2-way or 4-way tensors.

\begin{figure}[h]
    \centering
    \includegraphics[width=0.8\linewidth]{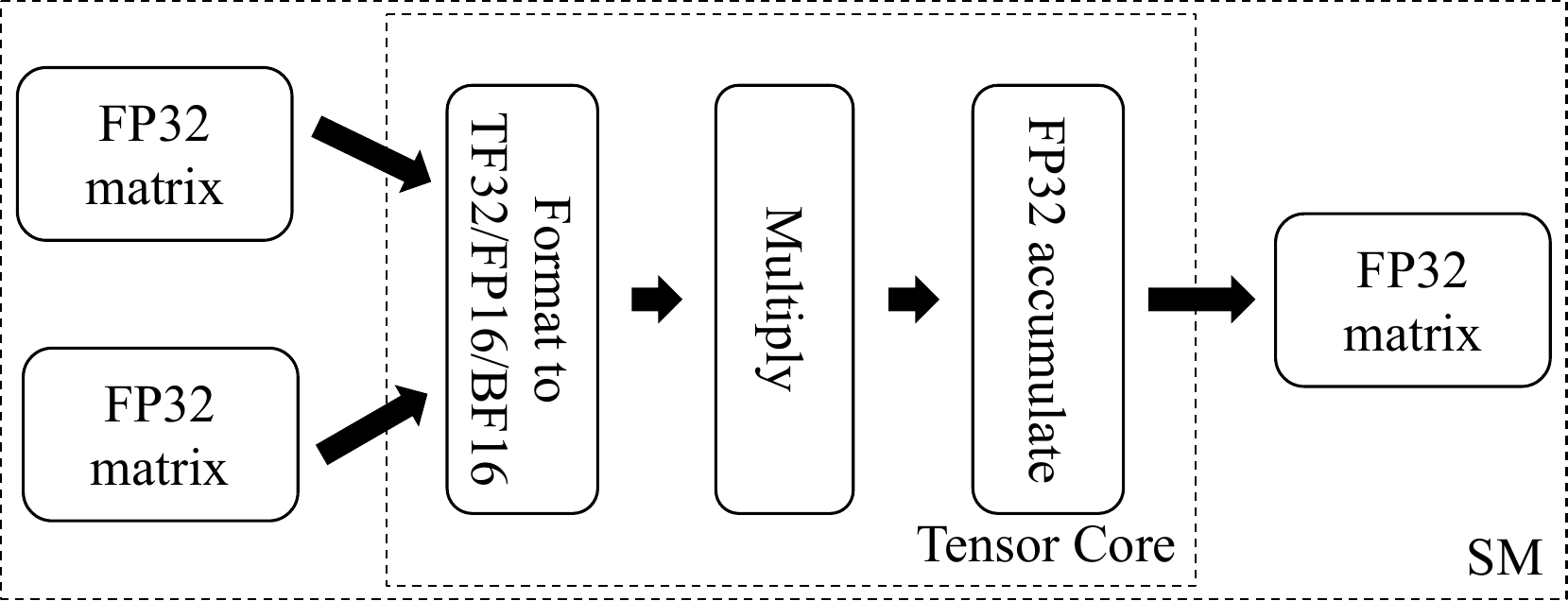}
    \caption{FP32 input/output matrices multiplication using the Tensor Core (source: NVIDIA).}
    \label{fig:chap2_tc_accumulator}
    % \Description{}
\end{figure}

Another issue to consider when using Tensor Cores is precision. While Tensor Cores support FP64 in Ampere and subsequent architectures, it is beneficial to explore lower precision compute modes such as TF32, FP16, and BF16 to attain higher performance. As shown in Table~\ref{tab:chap2_a100__format_perf}, the peak performance of TF32/FP16/BF16 is 8-16 times greater than that of FP32/FP64. To utilize the peak performance of TF32/FP16/BF16 format, Tensor Cores provide the function of converting high-precision inputs into low-precision formats during computation, and using single-precision format in the accumulation process and output format (see Figure~\ref{fig:chap2_tc_accumulator}). This can yield a highly favorable acceleration ratio while satisfying the precision requirements of applications such as artificial intelligence.  Nevertheless, precision issues may arise when employing TF32/FP16/BF16 in precision-sensitive scientific computing, as their narrower exponent range or fewer significant digits compared to FP32/FP64 may lead to imprecision, necessitating careful consideration and remediation~\cite{markidis2018nvidia, ootomo2022recovering, ootomo2023quantum}.

\subsection{Efficient tensor contraction on GPU} %% Henry
\label{section:contraction_on_gpu}
%% demonstrate two efficient implementation/algorithms of einsum contractions
Any dimensional tensor contraction can be expressed by a summation formula as follows:

\begin{equation}
\begin{aligned}
    C_{\gamma_{1}\gamma_{2}...\gamma_{d_C}}=\sum_{\delta_{1}}\sum_{\delta_{2}}...\sum_{\delta_{d_S}}{A_{\alpha_{1}\alpha_{2}...\alpha_{d_A}}B_{\beta_{1}\beta_{2}...\beta_{d_B}}}
\end{aligned}
\label{eq:chap2_einsum}
\end{equation}
where $\delta_{1}\delta_{2}...\delta_{d_S} = (\alpha_{1}\alpha_{2}...\alpha_{d_A})\cap(\beta_{1}\beta_{2}...\beta_{d_B})$ are the contracting indices for both tensors $A$ and $B$, which are summed, and 
\begin{displaymath}
\gamma_{1}\gamma_{2}...\gamma_{d_C} = (\alpha_{1}\alpha_{2}...\alpha_{d_A})\cup(\beta_{1}\beta_{2}...\beta_{d_B}) \backslash (\delta_{1}\delta_{2}...\delta_{d_S})
\end{displaymath}
are the remaining indices for both tensors $A$ and $B$. The dimensions of tensors $A$, $B$, and $C$ are $d_A$, $d_B$, and $d_C$, respectively and $d_S$ is the number of contracting indices.

The computing complexity ($T_{cc}$) of the tensor contraction
\begin{equation}
\begin{aligned}
    T_{cc}=\text{ops\_per\_element} * {\prod \limits_{i=1}^{d_A} \text{len}(\alpha_{i}) \prod \limits_{i=1}^{d_B} \text{len}(\beta_{i})}/{\prod \limits_{i=1}^{d_S} \text{len}(\delta_{i})},
\end{aligned}
\end{equation}
is the product of total dimensions of each tensor divided by the total dimension of the contracting indices. The term ops\_per\_element represents the number of floating-point operations required for the multiplication of elements, for example, $\text{ops\_per\_element} = 8$ for complex number multiplication.

The memory complexity ($T_{mc}$) for the contraction is the sum of the memory required to store the two input tensors A,B and the output tensor C
\begin{equation}
\begin{aligned}
    T_{mc}=\text{sizeof\_data}*(\prod \limits_{i=1}^{d_A} \text{len}(\alpha_{i}) + \prod \limits_{i=1}^{d_B} \text{len}(\beta_{i}) + \prod \limits_{i=1}^{d_C} \text{len}(\gamma_{i})).
\end{aligned}
\end{equation}
The term sizeof\_data represents the number of bytes for an element. One important indicator to assess whether tensor contractions can leverage the performance of a GPU platform is the arithmetic intensity, which is defined as the ratio of $T_{cc}$ and $T_{mc}$.

For high performance implementation of tensor contraction on GPU, the main difficulty lies in the arbitrariness of tensor dimension and tensor contraction indices. Generally speaking, there are two main ways to efficiently implement tensor contraction on GPU:\\
% \begin{itemize}
    \textbf{Transpose-Transpose GEMM Transpose (TTGT)} contraction: The TTGT method mainly divides tensor contraction into two types of operations: tensor transpose and matrix-matrix multiplication (GEMM). The purpose of tensor transpose is to reshape a tensor of arbitrary dimension into a flat two-dimensional matrix, where the uncontracting indices are merged into one dimension and the contracting indices are merged into another dimension and vice versa. In this way, the transposed tensor can be computed using the already highly optimized GEMM operation, and finally, the GEMM calculation result can be transposed back to the result tensor as needed. The advantage of this method is its versatility. The operation decomposition allows to utilize specialized transpose operation libraries (such as cuTT~\cite{hynninen2017cutt} and TTLG~\cite{vedurada2018ttlg}) and GEMM libraries (such as cuBLAS and CUTLASS) separately on the GPU. However, this method also has obvious drawbacks: 1) Tensor transpose requires additional temporary space to store the transposed tensor, which is fatal in quantum circuit simulation because of its high memory requirement; 2) The arithmetic intensity of TTGT will be much lower compared to an equally-sized GEMM because each element needs to be loaded at least twice.\\
    \textbf{GEMM-like Tensor-Tensor (GETT)} contraction~\cite{springer2019high}: The GETT method is based on a cache-based hierarchical and size-unrolled tensor contraction loop, which explicitly loads sub-loops into each level of GPU cache. This avoids the need for explicit preprocessing steps and allows for operations on sub-tensors of any order while preserving memory access with stride 1. The advantages of this method are: 1) No need for explicit tensor transposition, avoiding additional temporary storage overhead; 2) Can preserve arithmetic intensity for any given tensor contraction compared to the equally-sized GEMM. The disadvantage is that optimized routines need to be implemented separately for each individual tensor contraction case. The GPU-accelerated tensor library cuTensor uses this method~\cite{springer2019cutensor}.
% \end{itemize}

%\section{Methods}
%% Now this section is organized with two subsections Algorithm and Implementation
%% Or tell the story from the sides of Precision and Effiency as Heny suggested. 

\section{Simulation Algorithm}
\label{section:algorithm}
%%	• Algorithm  - Pan
%%		○ contraction path searching  
%%			§ cuTensor  opts (perf & larger intermediate tensor )
%%		○ How to contraction? sparse einsum?  (PRL 附录) -> 其他 复用缩并路径

In this section, we will present the overall simulation algorithm. The purpose of the simulation method is to accurately calculate the amplitudes of sampled bitstrings from quantum experiments. The simulation method consists of three main steps: contraction path finding, dynamic slicing, and sparse-state contraction, which will be discussed in detail below.

\subsection{Contraction path representations}
A contraction path is a sequential order of tensors that are to be contracted at each step.
In this study, we have limited our focus to the pairwise contraction approach, where two tensors are contracted at each step.
For pairwise contractions, the contraction path's length for a $N$-tensor network is $N-1$, where each position within the path specifies the indices of the tensors to be contracted.
The contracted tensor resulting from each step is indexed by the first tensor to be processed.

Alternative representations of the contraction path can be useful for instructional purposes. One such representation is the contraction tree~\cite{ogorman, gray2021Hyperoptimized}.
The contraction tree is a binary tree in which the leaf nodes signify the tensors to be contracted. The parent vertices of these leaf nodes correspond to the resulting tensor from each pairwise contraction operation.
Once all the tensors have been contracted, the leaves of the tree are merged into a single root vertex, which represents the final tensor. Figure~\ref{fig:simulation demo}(a) and (b) depicts a tensor network and a possible contraction tree to contract it.

\subsection{Path finding approach}
Finding a contraction path through a tensor network is a challenging combinatorial optimization problem that depends on the sequence of operations. It is an NP-hard problem, and to date, there is no efficient algorithm available to solve it.
However, significant progress has been made in recent years by leveraging graph partitioning tools to determine an optimal contraction order. \cite{gray2021Hyperoptimized} have demonstrated that this approach dramatically improves performance for a broad range of quantum circuits.
The state-of-the-art algorithm used to solve this problem mimics the simulated annealing (SA) process~\cite{kalachev2021Recursive}.
By implementing a SA-like approach and defining local transformations of the contraction tree, this problem can be resolved efficiently.

The SA-based algorithm used to solve the contraction path problem employs a score function that is analogous to an energy function in physics. As described in~\cite{kalachev2021Recursive}, the classical form of the score function is expressed, 
\begin{equation}
    \log(T_{cc} + \alpha T_{mc}) + \beta\log(T_{sc}),
\end{equation}
where $\alpha$ is a parameter that adjusts the balance between computing complexity and memory complexity according to the hardware. $T_{sc}$ denotes the size of the largest intermediate tensor, and $\beta$ is a parameter controlling the fraction of $T_{sc}$ in the score function.

\begin{figure}[thb]
    \centering
    \includegraphics[width=\linewidth]{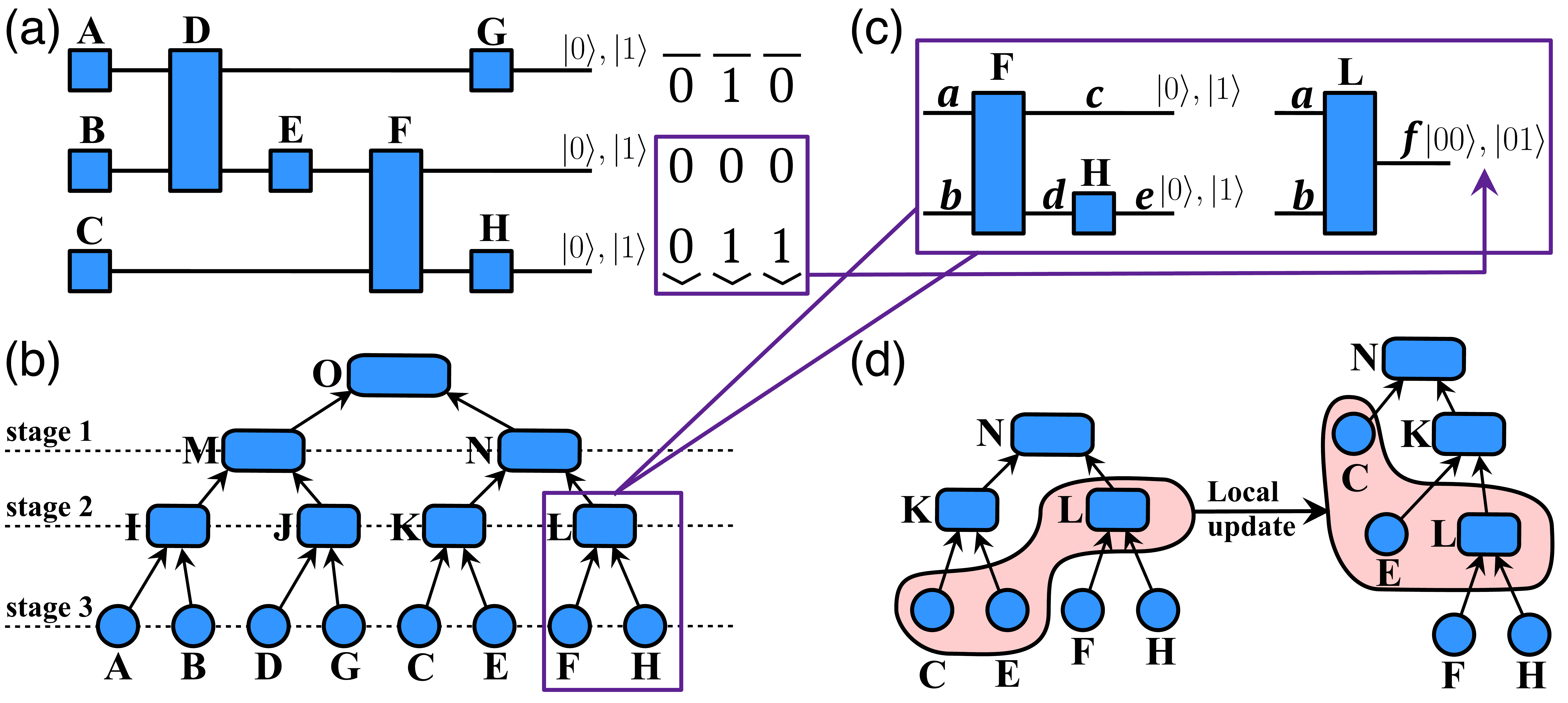}
    \caption{Overall scheme for the sparse state simulation method. (a) A tensor network that represent a quantum circuit, the aim of the simulation task is calculating the amplitudes of the sparse-state on the right-hand side of the network. (b) A contraction tree represent the unique contraction order of the tensor network in (a). (c) Sparse einsum operation, wherein two tensors with open bonds are contracted, and the merged bond $f$ is calculated based on the sparse-state in (a). (d) A local update that transforms contraction sub-tree from (b) to a new one. The local update area is marked by shaded red.}
    \label{fig:simulation demo}
    % \Description{}
\end{figure}

Once the score function for the contraction path finding task has been defined, two further components are necessary to implement an efficient SA-like optimization algorithm: local update and sweep.
In spin models, a local update typically involves a single spin flip, while a sweep involves traversing all spins in the system. However, in the context of contraction path finding, a local update is defined as exchanging descendants of the same ancestor vertex in a contraction tree. To restrict the exchange to be non-trivial and local, it is limited to three nearest descendants~\cite{kalachev2021Recursive}.  This constraint leads to two possible directions to move. An example of a local update in contraction trees is illustrated in Figure~\ref{fig:simulation demo}(d). Here, the original contraction order of three descendants ($C$, $E$, and $L$) is starting the contraction with the former two tensors followed by the contraction of the result with the third tensor. The local update changes the contraction order to contracting the last two tensors first and then contracting with tensor $C$.
As for sweeping in contraction trees, it refers to the process of traversing all vertices from the root to the leaves~\cite{Liu2022}.

% \subsection{Dynamic slicing}
The size of the largest intermediate tensor in arbitrary tensor network contraction is known to be exponential to the tree-width of the underlying graph~\cite{markov2008Simulating}. Consequently, executing the entire contraction at once on current computational hardware may be impossible. In such cases, it becomes necessary to slice the entire contraction into smaller sub-tasks that can be executed on available computational resources.
The sub-tasks must be designed to ensure they can be executed within the computational limitations of the hardware. During simulation, these sub-tasks are independent of each other and can be parallelly implemented on different computing devices. This process of parallelization substantially enables larger and more complex tensor networks to be effectively simulated on existing hardware.

The slicing technique is a relatively straightforward approach in tensor network simulation of quantum circuits. As summation indices are typically located at the connections between tensors, slicing involves cutting these connections and fragmenting the tensor network into $d$ smaller pieces, where $d$ is typically equal to 2 in the case of quantum circuits.
These smaller pieces have a slightly altered structure than the original network, potentially resulting in a change of the optimal contraction order~\cite{chen2018Classical}. Therefore, after each slicing step, the contraction order must be fine-tuned to fit the new tensor network structure, until the largest intermediate tensor can be contained within the computational hardware. This iterative slicing-fine tuning process is referred to as the dynamic slicing process in tensor network contraction by partitioning.
The dynamic slicing process is a crucial step in efficiently executing tensor network contractions on existing computational hardware, especially in the context of quantum circuit simulations.

\subsection{Sparse-state contraction}
\label{section:sparse_einsum}
Upon determining the contraction order and slicing scheme of the tensor network, the next step in simulation is to perform numerical contraction. Traditional tensor network contractions rely on vanilla einsum or tensordot operations. However, in the sparse-state simulation context, tensor network contractions differ somewhat and require the introduction of "sparse einsum" operations.

The primary distinction arises when two open bonds come together. As mentioned in the preceding section, open bonds in the tensor network simulation of quantum circuits epitomize the distinct layout of the final quantum states. Typically, the bonds of two tensors with differing open bonds are joined through reshape operations into a single entity, merging two indices into one. The dimension of conjoined indices corresponds precisely to the size of the Hilbert space that emerges from the tensor-product of the individual qubits' Hilbert spaces indicated by the open bonds before the merge. 

In our task, only bitstrings sampled from quantum experiments are required, making it unnecessary to retain the entire dimension of merged open bonds. To restore the required information from merged open bonds, we utilize bitstring samples to determine the entries to keep while contracting tensors. For instance, in Figure~\ref{fig:simulation demo}(c), when tensors $F$ and $H$ are contracted, $c$ and $e$, representing open bonds, are merged. To match the sparse-state, which only contains the unique configurations $|100\rangle$, $|101\rangle$, and $|001\rangle$, the dimension of merged bond $f$ should be 2, corresponding to $|00\rangle$ and $|01\rangle$. In terms of an equation, this operation can be expressed as
\begin{equation}
L_{fab} = \text{concat}(\sum_{d} F_{c=0,abd}H_{e=0,d}, \sum_{d} F_{c=0,abd}H_{e=1,d}),
\end{equation}
where the concatenation occurs in the dimension of $f$, and the subscripts of tensors represent their respective indices. This operation, referred to as "sparse einsum," is essential in simulating quantum circuits since the majority of states that arise in practical simulations are typically sparse.

% Since our task in this article is the information of neither a single bitstring nor the entire Hilbert space, but bitstrings sampled from quantum experiments, it is unnecessary to keep the entire dimension of merged open bonds. Due to the constant number of measurements in sampling based quantum experiments and the exponential size of Hilbert space, the sampled bitstrings compose a sparse state compared to the full quantum state. Therefore the number of entries holding the information we need in the merge open bonds tend to be small, especially for large numbers of qubits. To address the sparsity problem of merged open bonds, we will refer to the bitstring samples to determine which entries to keep when contracting tensors contain them. For example in Figure~\ref{fig:simulation demo}(c), when contracting $T_6$ and $T_8$, $a$ and $e$ are two open bonds to be merged. In order to match the sparse-state which contains only $|000\rangle$, $|101\rangle$ and $|001\rangle$, the dimension of merged bond $f$ will be 2, which correspond to $|00\rangle$ and $|01\rangle$ unique configurations of bond $a$ and $e$. In equation, this operation can be written as
% \begin{equation}
%     L_{fab} = \text{concat}\left(\sum\limits_{d} F_{c=0,abd}H_{e=0,d}, \sum\limits_{d} F_{c=0,abd}H_{e=1,d}\right)
% \end{equation}
% here concatenation occurs in the dimension of $f$, and the subscripts of tensors represent their indices. We dub this operation "sparse einsum".

By adhering to the restriction of the sparse state, composed of bitstring samples, its amplitudes can be obtained through a single tensor network contraction.

\section{Optimization on GPU}
\label{section:opt on GPU}
%%	• Implementation - Henry 
%%		○ Tensor Core extended precision (TF32 -> 3xTF32 -> 3xFP16/BF16) 
%%		○ Mixed precision (展开，正确性 说明, TF32+3xTF32 mixed ) - Lvlin + Daochen 
%%			§ 14-layer full simulation (FP64/FP32/TF32/3xTF32)
%%              Through digital bits #  
%%		○ einsum to GEMM 
%% 

%%

%% each part shoud include benchmark/comparison
\subsection{Convert Einsum into GEMM in tensor network} %% Henry
\label{section:einsum2gemm}
%% demonstrate why and how, conculde a strategy, focus on sparse-batched einsum conversion
%% Benchmark: perf
%% Benchmark example: isolate related contractions in 18-layer example
%% ○ Balanced index 
%% ○ 20-layer perform better 

% \begin{figure}[h]
%     \centering
%     \includegraphics[scale=0.6]{figs/chap4_contraction_flow.pdf}
%     \caption{A tensor contraction tree: the arrow denotes the data storage direction, where B->A represents A = einsum(A, B). The overall contraction direction is from right to left, top to bottom. The final state of the tensor network after contraction is stored in tensor A.}
%     \label{fig:chap4_contraction_flow}
%     \Description{}
% \end{figure}

In a contraction tree like Figure~\ref{fig:simulation demo}(b), according to the distance from the final tensor, the tensor contraction can be divided into stage 1, stage 2, etc., with the most time-consuming part concentrated in the stage 1, which is the process of contracting tensors to the stem tensor. As discussed in Section~\ref{section:sparse_einsum}, the tensor contraction process of the tensor network involves a large number of einsum, batch einsum, and sparse einsum. Therefore, optimizing the contraction of these tensors is crucial for improving the efficiency of the entire tensor network contraction.

Generally speaking, the tensor network can contain hundreds or thousands of tensors. Determining the contraction order and the index arrangement of each contraction is an NP-hard problem. The index arrangement has a significant impact on the efficiency of the computation on the GPU. As mentioned in Section~\ref{section:contraction_on_gpu}, when the contracting indices and remaining indices are located in the front and rear parts of the tensor indices after transposition, respectively, the tensor contraction can be represented as a GEMM operation. Therefore, after the tensor contraction path is determined, we can perform index reordering on certain einsums along the critical path to obtain performance benefits at a relatively small cost. 

\begin{figure}[h]
    \centering
    \includegraphics[width=\linewidth]{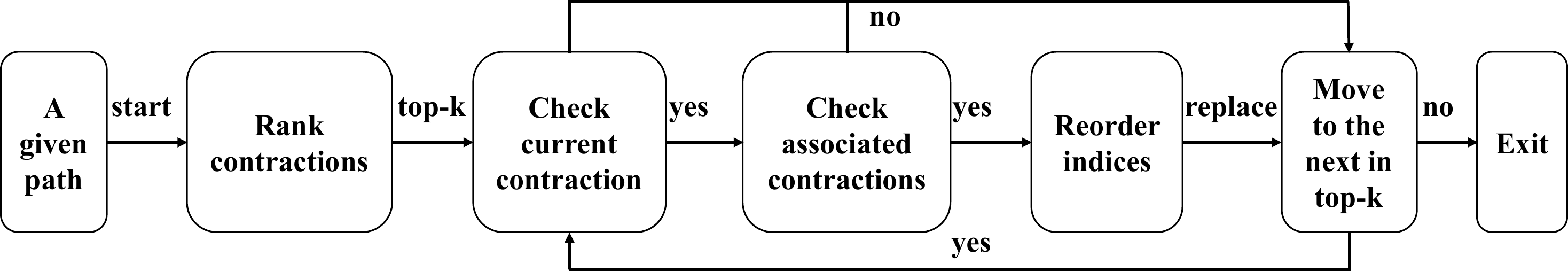}
    \caption{Reorder indices for top-k contractions in a tensor network.}
    \label{fig:chap4_reordering_eqs}
    % \Description{}
\end{figure}

Figure~\ref{fig:chap4_reordering_eqs} shows the index reordering process for the determined path, which mainly includes the following parts: \\
% \begin{itemize}
    \textbf{Rank contractions}: Rank all contractions based on computing complexity or actual time consumption, and obtain the top-k contractions group with the highest computing complexity or time consumption. \\
   \textbf{Check current contraction}: Starting from the most time-consuming contraction, check whether it is suitable for reordering indices. The criteria for reordering indices are: 1) it has not been modified by other reordering steps; 2) only by modifying the input index order can it form a GEMM expression without modifying the output order of the current contraction. If both are possible, proceed to the next step to check the associated contraction. If not, move on to the next contraction in the top-k group. \\
   \textbf{Check associated contractions}: Here, the associated contractions refer to the contractions related to the two input tensors from the previous step. For example, as shown in Figure~\ref{fig:simulation demo}(b), if the current contraction O = einsum(M, N) is determined to reorder, then check if the output indices of the associated contractions for M = einsum(I, J) and N = einsum(K, L) can be reordered. The criteria for this are that neither of these two contractions has been modified by other reordering steps. If reordering is possible, proceed to reorder the output indices of the relevant contractions. If not, move on to the next contraction in the top-k group.\\
   \textbf{Reorder indices}: If both previous two steps are confirmed to be reordered, then in this step, reorder and replace the indices of the current contraction and associated contractions in the top-k group.
% \end{itemize}

% \begin{figure}[h]
%     \centering
%     \includegraphics[scale=0.7]{figs/chap4_reorder_eqs.pdf}
%     \caption{Reorder equations for a given path in a tensor network.}
%     \label{fig:chap4_reordering_eqs}
%     \Description{}
% \end{figure}

% \begin{figure}[h]
%     \centering
%     \includegraphics[width=\linewidth]{figs/chap4_top10_comparison_20layer_0.pdf}
%     \caption{Top 10 time-consuming equations in 18-cycle and 20-cycle circuit.}
%     \label{fig:chap4_top10_eqs}
%     \Description{The ranking is based on the computing time on A100-80GB using 3$\times$TF32.}
% \end{figure}

\begin{figure}[h]
    \centering
    \includegraphics[width=0.8\linewidth]{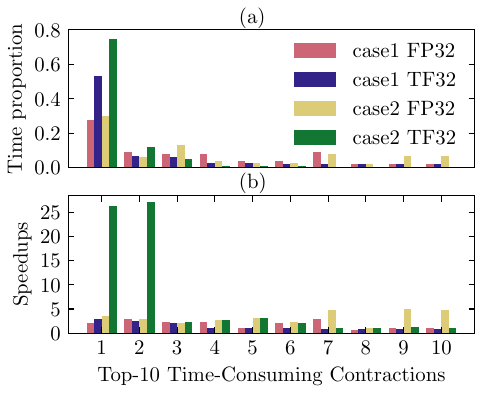}
    \caption{The time proportion and speedups of converting einsum to GEMM for the top-10 time-consuming contractions in two tensor network contraction cases.}
    \label{fig:chap4_top10_eqs}
    % \Description{The ranking is based on the computing time on A100-80GB using TF32.}
\end{figure}

%% + compare to torch.bmm 
Overall, the index reordering for the top-k contractions is similar to the transpose process in the previously mentioned TTGT algorithm. The difference is that instead of directly transposing tensors at one step, the indices of the output tensors are reordered according to the order required by the next contractions, thus avoiding additional temporary storage overhead. The cost can be ignored compared to the transposition in TTGT algorithm since it only affects the output of two input tensors. After reordering the indices of the top-k contractions, we can usually make the most time-consuming contractions in the contraction tree a GEMM operation, thereby improving the GPU execution efficiency of the critical contraction. Figure~\ref{fig:chap4_top10_eqs} shows two examples of tensor network contractions (i.e., experiments in the Section~\ref{section:experiments}) computed using FP32 and TF32. Figure~\ref{fig:chap4_top10_eqs}(a) displays the percentage of time spent on the top-10 contractions, which are shown to be the most time-consuming ones. It is evident that the computation is concentrated on a few contractions. Figure~\ref{fig:chap4_top10_eqs}(b) shows the speedups of transforming the top-10 contractions from Einsum to GEMM, which ranges from 1x to 26x.

Furthermore, as mentioned in Section~\ref{section:sparse_einsum}, the tensor network contraction for the sampling problem includes sparse einsum operations, which are often the most time-consuming part of the entire contraction tree contraction process. To the authors' knowledge, this operation is not supported natively by any existing computing library to this date. A compromise implementation is to perform slicing first, extract the required tensors from the original tensor for aggregation, and then perform batched einsum calculation. This method has two main drawbacks: 1) slicing increases the amount of data to be read, and the slicing time may become the bottleneck of the entire sparse einsum calculation; 2) it increases the memory requirement for storing intermediate sliced tensors, which is not friendly to quantum circuit simulation with limited memory resources. In this routine, after the above reordering step, since the \textit{cublasGemmBatchedEx()} API in cuBLAS supports passing separate pointers for each matrix of the batched GEMM input, we can directly convert the reordered sparse einsum into batched GEMM. This completely eliminates the slicing process for computing sparse einsum. In Figure~\ref{fig:chap4_top10_eqs}, the two most time-consuming contractions are both sparse einsums. It can be seen that after reordering and using batched GEMM, their calculation efficiency can be improved up to 26 times.

As quantum circuits become deeper and more complex, tensor contractions become denser, making the optimization strategy more effective.

\subsection{Extended precision} %% Henry, Petrick
%% demonstrate how 3xtf32, 3xfp16/bf16 work, start from the existing solution 3xtf32 in cutlass
%% Benchmark: accuracy and perf
%% Benchmark examples: different shape of einsum/gemm

%% Goal: to take advantage of the NVIDIA Tensor Cores to reach maximum performance.

% On NVIDIA A100, as shown in Table~\ref{tab:chap2_a100__format_perf}, the throughput of Tensor Core with TF32 precision is 8x faster than the FP32 CUDA Core math and Tensor Core with FP64 precision. The math throughput from Tensor Core with FP16/BF16 is 2x speedup compared to Tensor Core with TF32 precision, 16x speedup compared to FP32 CUDA Core math.

% In actual application scenarios, under the condition of satisfying the maximum error acceptable to the application, the Tensor Core instruction with the lowest possible data precision is usually selected to maximize the acceleration of matrix calculation. For example, for matrix calculations with FP32 data precision, using TF32 precision Tensor Core instructions can achieve nearly 8 times faster than using FP32 CUDA Core calculations with a loss of certain precision. 
% If you use FP16 precision Tensor Core instructions, you will get close to 16 times the speedup, but the cost is that the calculation accuracy will be further reduced.

In scientific computing, including quantum circuit simulations, significant decrease in calculation accuracy is usually unacceptable. As a result, the use of low-precision Tensor Cores to accelerate computations directly is often not feasible, despite their high throughput capabilities. In order to address this issue, CUTLASS introduced a new implementation in version 2.8 using three TF32 Tensor Core instructions to emulate the FP32 operations~\cite{nvcutlass_3xtf32}. Compared to native FP32 operations, the 3xTF32 method achieves between 2-3x speedup while maintaining the accuracy.

% In order to address this issue, version 2.8 of CUTLASS introduced a new implementation known as the 3$\times$TF32 method~\cite{nvcutlass_3xtf32}. This method uses three TF32 Tensor Core instructions to emulate the FP32 computation process while ensuring high accuracy and achieving a speedup of 2$\sim$3 times that of FP32 computations.
\begin{equation}
\begin{aligned}
    % A &= \text{tfloat32\_t}(A_{\text{big}}) + \text{tfloat32\_t}(A_{\text{small}}) \\
    % B &= \text{tfloat32\_t}(B_{\text{big}}) + \text{tfloat32\_t}(B_{\text{small}}) \\
    X_{\text{big}} &= \text{float\_to\_tf32}(X), \\
    X_{\text{small}} &= \text{float\_to\_tf32}((X - \text{float}(X_{\text{big}})), \\
    A * B + C &= (A_{\text{big}} + A_{\text{small}}) * (B_{\text{big}} + B_{\text{small}}) + C\\
    &\approx A_{\text{big}} * B_{\text{small}} + A_{\text{small}} * B_{\text{big}} + A_{\text{big}} * B_{\text{big}} + C.
\end{aligned}
\label{eq:3xTF32}
\end{equation}

The 3$\times$TF32 method is shown in Equation~\ref{eq:3xTF32}. The FP32 matrix elements of matrices A and B are split into large and small parts which are represented as TF32 data. For A matrix element a, the large part ${a_{\text{big}}}$ is the direct lossy conversion from the FP32 data to TF32, while the small part ${a_{\text{small}}}$ is residual of the conversion from a to ${a_{\text{big}}}$. In this way, the original multiplication and addition operation is split into four multiplication and addition operations. The operation of two residuals can usually be ignored, so the calculation is approximated as three TF32 Tensor Core operations. The use of 3$\times$TF32 expands the significant digits of TF32, and obtains significant digits similar to FP32, thereby improving the calculation accuracy. From the perspective of speedup, using 1$\times$TF32 can obtain 8x speedup compared to FP32, while using 3$\times$TF32 can theoretically obtain 8/3, which is close to 2.3x speedup.

 % Among them, ${A_{\text{big}}}$ is directly converted from element A using TF32 data type, and the decimal is ${A_{\text{small}}}$, which represents the residual of A and ${A_{\text{big}}}$ using TF32. The elements of matrix B are treated in the same way.

We extend the above method to FP16 and BF16 precision to implement the 3$\times$FP16/3$\times$BF16 method. As shown in Table~\ref{tab:chap2_a100__format_perf}, the precision of the FP16 data type is equivalent to the TF32, but the exponent part can only represent data with a small dynamic range. The BF16 data type has the same dynamic range as the TF32 data type, but lower data precision. Therefore, if the actual data range of the application is within the dynamic range of FP16, 3$\times$FP16 can obtain the same calculation accuracy as 3$\times$TF32, but at a 2x speedup theoretically. When the data exceeds the dynamic range of FP16, we can also consider using 3$\times$BF16 to obtain a trade-off between speedup and accuracy.

We extended CUTLASS's support to 3$\times$FP16 and 3$\times$BF16 following the 3$\times$TF32 method. To simplify the complexity of precision testing, we use vector dot products to study the effects of different methods on the resulting accuracy. This is because both einsum and GEMM can be seen as composed of non-interfering basic vector dot products. We use the data range from 1e3 to 1e-7, which we refer to as FP16 range later in the remainder of this article, to test the calculation accuracy of different calculation methods.

% We use two data ranges to test the calculation accuracy of different calculation methods.  The range from 1e3 to 1e-7 ,which will be called FP16 range later, is within the dynamic range of FP16. The range from 1e-2 to 1e-12 ,which will be called QC(Quantum Computing) range, is a common data range in quantum computing, and part of this range has exceeded the dynamic range of FP16.

\begin{figure}[h]
  \centering
  \includegraphics[width=0.7\linewidth]{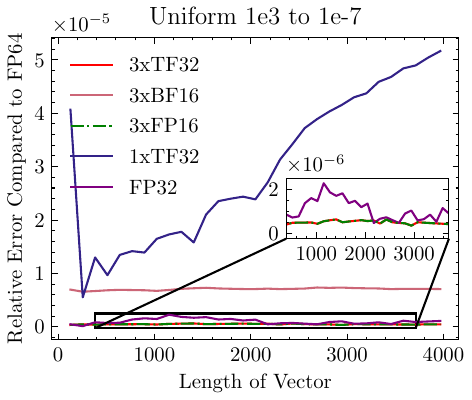}
  \caption{Relative Error comparison of FP32, 1$\times$TF32, 3$\times$FP16, 3$\times$BF16, 3$\times$TF32 on 1e+3 $\sim$ 1e-7.}
  \label{fig:1e+3_1e-7_1xTF32}
\end{figure}

Figure~\ref{fig:1e+3_1e-7_1xTF32} shows the relative error comparison of 1$\times$TF32, 3$\times$BF16, 3$\times$TF32, 3$\times$FP16 and FP32 under FP16 range. It can be concluded that the errors calculated by 3$\times$TF32 and FP32 are roughly the same, the error of 1$\times$TF32 is significantly larger than that of 3$\times$TF32, and the error of 3$\times$BF16 is between the two. According to the previous discussion, under the FP16 range, the 3$\times$FP16 method should obtain the same calculation accuracy as 3$\times$TF32. The zoomed view in Figure~\ref{fig:1e+3_1e-7_1xTF32} shows that the errors of 3$\times$TF32 and 3$\times$FP16 are exactly the same. They are even better than the calculation accuracy of FP32 in this test. However, 3$\times$BF16 is limited by the significant digits that BF16 can represent, making its calculation error relatively larger, but still smaller than 1$\times$TF32.

% \begin{figure}[h]
%   \centering
%   \includegraphics[width=\linewidth]{figs/log_128__4096_1e-2_1e-12_all_compared.pdf}
%   \caption{Relative Error comparison of FP32, 1$\times$TF32, 3$\times$BF16, 3$\times$TF32 on 1e-2 $\sim$ 1e-12.}
%   \label{fig:1e-2_1e-12_1xTF32}
% \end{figure}

% Figure~\ref{fig:1e-2_1e-12_1xTF32} shows the calculation errors of different methods under the QC range. Same as the above conclusion, 3$\times$TF32 can still obtain calculation accuracy equivalent to FP32, the error of 1$\times$TF32 is still relatively large, and the calculation error of 3$\times$BF16 is between 3$\times$TF32 and 1$\times$TF32. Because the dynamic range of QC range exceeds the dynamic range of FP16, the data beyond the range becomes 0, resulting in a significant drop in the calculation accuracy of 3$\times$FP16, even worse than 3$\times$BF16. And if the dynamic range of the data exceeds the maximum value of FP16, the calculation error of 3$\times$FP16 will be significantly larger.

Through the above experiments, it can be concluded that the 3$\times$TF32 method can obtain calculation accuracy equivalent to that of FP32. If the data range satisfies the FP16 dynamic range, we can directly use the 3$\times$FP16 method to obtain further acceleration. When the data accuracy exceeds the FP16 dynamic range, the 3$\times$BF16 method serves as a compromise between performance and calculation accuracy. Additionally, projecting data onto the FP16 data representation range through dynamic scaling can also be an option.

\subsection{Mixed precision} %% Lvlin, Daochen, Henry
%% mixed is confusing, new name
%% demonstrate how tf32+3xtf32 work, how to compare results in a circuit simulation(norm2, xeb)
%% Benchmark: accuracy and perf
%% Benchmark example: 14-layer circuit (add tests for more configurations of mixed ratio, discuss with Henry)

% Need to explain why we only test 3xTF32 here, not including 3xBF16 (maybe we could say it can be added before August)

% Table - Smaller circuit results with FP64, FP32, TF32 and 3xTF32 precision.

% Single Tensor Core instruction (TF32/FP16/BF16) can provide high floating-point peak performance, but with lower precision. Multiple Tensor Core instructions (3$\times$TF32/FP16/BF16) can provide higher precision, but the peak performance theoretically only reaches one-third or less of a single Tensor Core instruction. Quantum circuit simulation has different precision requirements for different types of computing tasks. Therefore, according to the requirements of the specific computing task, it is necessary to explore the balance between precision and performance to maximize the computing power of GPU Tensor Core.

A single Tensor Core instruction (TF32/FP16/BF16) can offer high peak performance for floating-point calculations, albeit with lower precision. On the other hand, multiple Tensor Core instructions (3$\times$TF32/FP16/BF16) can provide higher precision, but the peak performance may only reach up to one-third or less of a single Tensor Core instruction. The precision requirements of quantum circuit simulation vary for different types of computing tasks. Therefore, it is crucial to explore the balance between precision and performance according to the specific computing task's requirements to maximize the computing power of the GPU's Tensor Cores.

\begin{scriptsize}
\begin{table}
\centering
  \caption{Comparison of $F_{l}$ and time for 45-qubit 14-cycle circuit with different precision}
  \label{tab:n45m14_xeb}
  \begin{tabular}{llll}
    \hline
    Precision & $F_{l}$ & $\epsilon_{F_{l}}$(\%) & Speedups \\
    \hline
    FP64 & \textbf{0.004753936} & 0.0 & 1 \\
    FP32 & \textbf{0.0047539}47 & 2.28E-4 & 1.49 \\
    3$\times$TF32 & \textbf{0.0047}30105 & 0.5 & 2.57 \\
    TF32 & \textbf{0.004}288435  & 9.79 & 3.22 \\
    \hline
\end{tabular}
\end{table}
\end{scriptsize}

In this section, we utilized the $F_{l}$ verification problem of a relatively small-scale Sycamore quantum circuit instance with 45-qubit and 14-cycle to show that by adjusting the computation precision of different parts in the tensor network simulation it is possible to achieve a balance between computation efficiency and accuracy. The problem involves calculating amplitudes for $2^{20}$ sampled bitstrings and contains $2^{14}$ sub-tasks after slicing, where each sub-task performs 280 contraction steps. We performed the tensor network contraction using various precision formats, namely FP64, FP32, 3$\times$TF32, and TF32, to calculate the $F_{l}$ results at these precisions. Table~\ref{tab:n45m14_xeb} presents the impact of different precisions on the accuracy and performance of the final result. As expected, the performance improves with increasing precision while the accuracy decreases. Notably, the result of 3$\times$TF32 is acceptable, with only a 0.5\% difference from the FP64 result. However, the TF32 result is clearly unacceptable, with a 9.79\% error. Hence, we investigated the effects of different ratios of TF32 and 3$\times$TF32 on accuracy and performance by adjusting the ratio based on the computing complexity in tensor network contraction calculation. We did not adopt 3$\times$FP16/BF16 because they are not supported by the latest version of cuTensor.

\begin{figure}[h]
  \centering
  \includegraphics[width=0.85\linewidth]{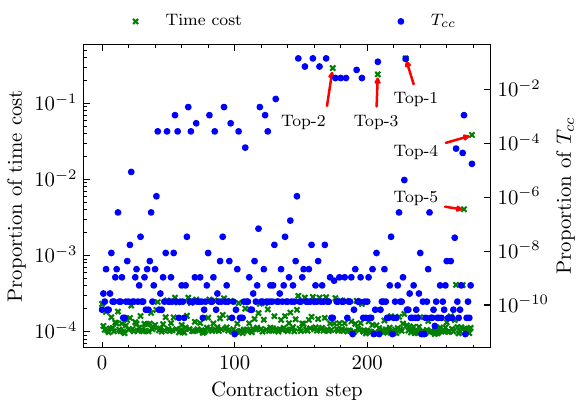}
  \caption{Time cost and computing complexity for each contraction step, the top-5 most time-consuming contractions are highlighted.}
  \label{fig:n45m14_Time-CC-step}
\end{figure}

Based on the characteristics of tensor network contraction, the computation is concentrated in the later stages of the contraction process, as shown in Figure~\ref{fig:n45m14_Time-CC-step}. In order to investigate the impact of different TF32 ratios on computation efficiency, we first sorted all contractions according to the actual computation time, and then gradually converted the computation precision from 3$\times$TF32 to TF32 starting from the most time-consuming contraction. We used $\epsilon_{L2^2}$, which is the relative error for the squared L2-norm of the resulting tensor 

\begin{equation}
\begin{aligned}
\label{eq:l2_squared_error}
\epsilon_{{\text{L2}^2}} = {\lvert {\text{L2}^2} - {\text{L2}^2}^{\prime} \rvert}/{{\text{L2}^2}},\\
\end{aligned}
\end{equation}
to quantitatively analyze the effect of different TF32 ratios on the result accuracy. The squared L2-norm  was used since Equation~\ref{eq:lxeb_2} shows that $p_U(x_i)$ calculates the probability of a complex vector $x$, and can be expressed as the square of the L2-norm of $x$. Thus, Equation~\ref{eq:lxeb_2} can be written as:
\begin{equation}
\label{eq:lxeb_as_l2}
F_l = 2^N\sum\limits_{i=1}^mp_U(x_i)/m - 1 = {\text{L2}(x)}^2 2^N/m - 1.
\end{equation}

\begin{figure}[h]
 \centering
 \includegraphics[width=0.8\linewidth]{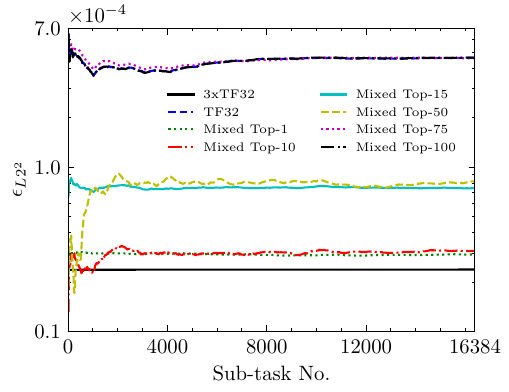}
 \caption{The relative errors of the squared L2-norm for different mixed.}
 \label{fig:n45m14_al2norm_err}
\end{figure}

%Figure~\ref{fig:n45m14_al2norm_err} shows the influence of different TF32 ratios on $\epsilon_{L2^2}$. It can be observed that as the replacement ratio increases, $\epsilon_{L2^2}$ also gradually increases. Compared with the 3$\times$TF32 computation, after completely replacing it with TF32, $\epsilon_{L2^2}$ increased by 20 times. At the same time, it can be seen that for a certain TF32 ratio, as the number of sub-tasks increases, $\epsilon_{L2^2}$ shows a stable trend. It is easy to draw the conclusion that once $\epsilon_{L2^2}$ tends to be stable, the final error of $F_{l}$ can be determined, because the relative error $\epsilon_{F_{l}}$ of $F_{l}$ can be expressed as $\epsilon_{L2^2}$:
Figure~\ref{fig:n45m14_al2norm_err} illustrates the impact of different TF32 ratios on $\epsilon_{L2^2}$. It is noticeable that as the replacement ratio increases, $\epsilon_{L2^2}$ also increases gradually. After replacing 3$\times$TF32 computation entirely with TF32, $\epsilon_{L2^2}$ increased by 20 times. Additionally, it can be observed that for a particular TF32 ratio, as the number of sub-tasks increases, $\epsilon_{L2^2}$ displays a stable trend. Consequently, once $\epsilon_{L2^2}$ becomes stable, the final error of $F_{l}$ can be determined because the relative error $\epsilon_{F_{l}}$ of $F_{l}$ can be expressed as $\epsilon_{L2^2}$:
\begin{equation}
\begin{aligned}
\label{eq:lxeb_as_l2_error}
\epsilon_{F_{l}} = {\lvert F_l - F_l^{\prime} \rvert}/{F_l} = \epsilon_{{\text{L2}^2}}{(1 + F_l)}/{F_l}.
\end{aligned}
\end{equation}

\begin{table}
\centering
  \caption{Accumulated $T_{cc}$ ratio replaced, relative time to that with TF32, and relative error after replacing precision of the top-k contractions from 3$\times$TF32 to TF32. }
  \label{tab:n45m14_time-error-cc}
  \begin{tabular}{llll}
    \hline
    Top-k & $T_{cc}$ ratio & Relative time & $\epsilon_{{\text{L2}^2}}$ \\
    \hline
    0   & 0 & 1.34  & $2.38\times 10^{-5}$  \\
    1   & 0.14 & 1.26  & $2.94\times 10^{-5}$  \\
    10  & 0.68 & 1.10  & $3.08\times 10^{-5}$  \\
    % 15  & 0.85 & 1.05  & $7.45\times 10^{-5}$  \\
    50  & $\sim~1$ & $\sim~1.00$  & $8.20\times 10^{-5}$  \\
    % 75  & $\sim~1$ & $\sim~1.00$  & $4.59\times 10^{-4}$ \\
    % 100 & $\sim~1$ & $\sim~1.00$  & $4.64\times 10^{-4}$ \\
    280 & 1 & 1 & $4.63\times 10^{-4}$  \\
    \hline
\end{tabular}
\end{table}
Table~\ref{tab:n45m14_time-error-cc} presents the impact of varying TF32 ratios on computation time and $\epsilon_{L2^2}$, showing that the error increases as more contractions use TF32. Notably, the calculation is concentrated on a few contractions, and reducing the precision of key contractions from 3$\times$TF32 to TF32 can considerably decrease computation time while keeping the overall accuracy within an acceptable range. For instance, replacing the top-10 contractions with TF32 reduces the relative TF32 computation time from 1.34 to 1.1, with a minor increase in $\epsilon_{L2^2}$ from $2.38\times10^{-5}$ to $3.08\times10^{-5}$.

% Table~\ref{tab:n45m14_time-error-cc} lists the impact of replacing different TF32 ratios on computation time and $\epsilon_{L2^2}$. The more contractions using TF32, the closer the error is to using all TF32. It can be seen that the calculation is concentrated on a few contractions. Replacing the precision of some key contractions from 3$\times$TF32 with TF32 can significantly reduce calculation time while maintaining the overall calculation accuracy within an acceptable range, e.g., when the top-10 contractions are replaced with TF32, the relative TF32 computation time decreases from 1.34 to 1.1, and at the same time, $\epsilon_{L2^2}$ only increases from $2.38\times10^{-5}$ to $3.08\times10^{-5}$.

% From the above example, it is concluded that we can achieve a balance between computation efficiency and accuracy by adjusting the computation precision of different parts in the tensor network.

\section{Experiments}
\label{section:experiments}
%% • Demonstration /Results / 
%%		○ 18-layer 
%%			§ results & analysis (performance)  - Pan
%%			§ precision: 18-layer partial simulation -> trend to estimated - Lvlin 
%%		○ demo 20-layer: compare to Sunway, energy efficiency (wait some results) - Pan/Henry/lvlin 
%%			§ TC,  30 compare to 35 ()
%%			§ Efficiency 1%
%%			§ Power   
In this section, we present a demonstration of the effectiveness of the sparse-state tensor network simulation method in both verification and bounded LXEB fidelity simulation tasks of Google's RCS experiment~\cite{arute2019Quantum} using the optimization methods mentioned previously. The circuits and bitstring samples employed in this study were retrieved from~\cite{sycamore_data}. Our experimental work was carried out on a DGX-A100~\cite{nvdgxa100} equipped with eight NVIDIA A100-SXM-80GB GPUs. The experiment used CUDA version 11.8, cuTensor version 1.6 with 3$\times$TF32 support, and cuBLAS version 11.11. Each A100 GPU executed partial sub-tasks independently, and the final outcome was the sum of the resulting tensors.

\subsection{18-cycle verification task}
%% Google provide 2.5M samples, 
%% we use 1M (40%) samples. 
%% 18-layer task description. 

To limit computational resource requirements to a reasonable size, we calculated amplitudes for $2^{20}$ out of 2.5 million bitstrings sampled from 53-qubit 18-cycle circuits, the estimated LXEB fidelity for all sampled bitstrings is $0.356\%$.

%% Error estimated value
% \begin{figure}[h]
%   \centering
%   \includegraphics[width=0.7\linewidth]{figs/n53m18_al2norm2_err.pdf}
%   \caption{Relative Errors of L2-norm squared with 3$\times$TF32 and mixed precision (3$\times$TF32+TF32)}
%   \label{fig:n53m18_al2norm2_err}
% \end{figure}

\begin{figure}[thb]
  \centering
  \includegraphics[width=\linewidth]{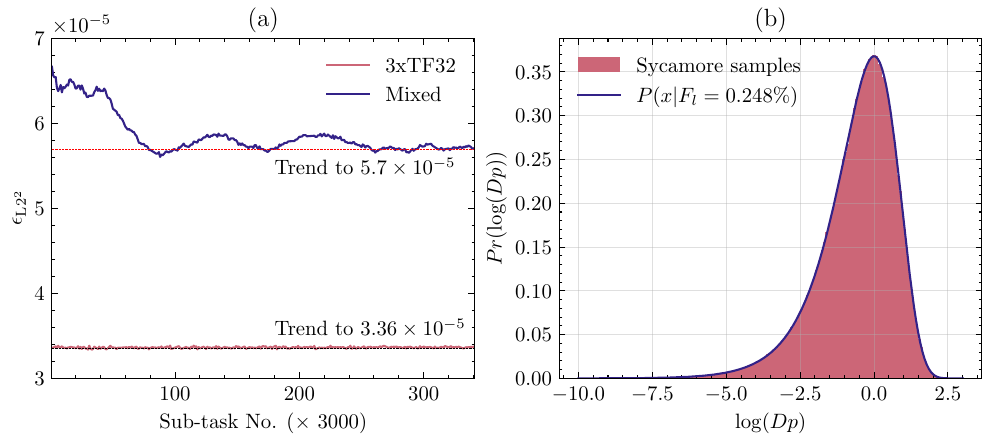}
  \caption{(a) Relative Errors of squared L2-norm with 3$\times$TF32 and mixed precision. (b) Histogram of the bitstring probability distribution over $2^{20}$ bitstring samples of 53-qubit 18-cycle Sycamore circuit. The purple line in (b) represents the theoretical distribution with 0.248\% LXEB fidelity.}% of $\log(Dp)$.}
  \label{fig:bitstring_dis}
\end{figure}

In the simulation optimization of the 18-cycle circuit, we first checked the top-10 contractions based on the principles outlined in Section~\ref{section:einsum2gemm}. The total computing complexity of these top-10 contractions accounts for 70.3\% of the entire tensor network computing complexity. After checking, a total of 50.62\% of the computing complexity can be converted from einsum into GEMM. In order to further improve the computational efficiency of the GEMM part, we also applied the mixed precision method to change the computational precision of the GEMM part from 3$\times$TF32 to TF32. For comparison, we calculated 1 million sub-tasks using FP32, 3$\times$TF32, and mixed methods, and then compared the relative errors of L2-norm squared. As shown in Figure~\ref{fig:bitstring_dis}(a), it can be seen that the $\epsilon_{{\text{L2}^2}}$ of 3xTF32 and mixed methods are stable with respect to the results of FP32, and are $3.36\times10^{-5}$ and $5.7\times10^{-5}$, respectively.

%% FP32, 3xTF32, mixed -> min time performance
%% Only use Table to list the time data
\begin{table}
\centering
  \caption{Computing time and sustained FLOPS with/without optimization and on different precisions on 53-qubit 18-cycle Sycamore RCS experiments verification task.}
  \label{tab:n53m18_time}
  \begin{tabular}{llll}
    \hline
    Precision/Optimization &  FP32/No & FP32/Yes & Mixed/Yes\\
    \hline
    Time per sub-task (sec.) & 14.32 & 9.99 & 3.62 \\
    \hline
    Time all tasks (GPU hours) & 33,368 & 23,278 & 8,435\\
    \hline
    Sustained FLOPS (TFLOPS) & 5.46 & 7.82 & 21.57\\
    \hline
\end{tabular}
\end{table}

%% Analysis
By implementing the aforementioned optimizations, we successfully calculated exact bitstring amplitudes for $2^{20}$ in 8435 GPU hours. Table~\ref{tab:n53m18_time} displays that the final optimized version with mixed precision yielded a 3.96-fold acceleration as compared to the non-optimized FP32 version. This acceleration is attributable to two factors: firstly, the optimization of converting sparse einsum to a pure FP32 GEMM operation allowed for a $30\%$ reduction in computational time, and secondly, the utilization of mixed precision resulted in a time reduction factor of 2.76.

% \begin{figure}[thb]
%   \centering
%   \includegraphics[width=0.85\linewidth]{figs/distribution.pdf}
%   \caption{Histogram of the bitstring probability distribution over $2^{20}$ bitstring samples of 53-qubit 18-cycle Sycamore circuit. The purple line represents the theoretical distribution.}% of $\log(Dp)$.}
%   \label{fig:bitstring_dis}
% \end{figure}

We computed the LXEB fidelity of $2^{20}$ bitstrings using Equation~\ref{eq:lxeb_2}. The resulting value was $F_{l} = (0.248\pm0.00571)\%$, where the error was estimated using Equation~\ref{eq:lxeb_as_l2_error}. Figure~\ref{fig:bitstring_dis}(b) illustrates the corresponding histogram of the bitstring probabilities, while the theoretical probabilistic density function~\cite{arute2019Quantum} of the sampled bitstrings is
\begin{equation}
    P(x\vert F_l) = (1+F_l(e^x - 1))e^{x-e^{x}},
\end{equation}
where $x \equiv \log(Dp)$ and $D=2^N$. The consistency between the numerical experiments and theoretical distribution indicates that the $2^{20}$ bitstrings of the 53-qubit 18-cycle do follow the Porter-Thomas distribution with estimated LXEB fidelity. While our calculated LXEB value does not exactly match Google's estimate, it is important to note that we only utilized approximately $42\%$ of the entire sample and there were ten circuit instances for the 18-cycle case. Thus, from a statistical point of view, our results provide validation for the output of Sycamore circuits with 53 qubits and 18 cycles.

%% Figure of bitstring distribution

\subsection{20-cycle verification and simulation task}
In order to compare our results with other simulation methods and implementations on different platforms, we will also examine the 20-cycle case. Considering the non-interacting sub-tasks in tensor network simulation, it is relatively straightforward to estimate the running time of a small fraction of sub-tasks and extrapolate it to the entire simulation.

% We employed our algorithm to estimate the computational cost and time required to compute all 3 million sampled bitstrings from the 53-qubit 20-cycle Sycamore circuits. This task had previously been accomplished using a CPU-based method~\cite{liu2022Validating}, which employed the entire supercomputer consisting of 41,932,800 cores with an overall single-precision performance of 1.5 EFLOPS. The final verification task was completed within 203 hours with a sustained performance of 84.8 PFLOPS. A summary of our verification details is presented in Table~\ref{tab:n53m20_val_time}. If we were to utilize the same single-precision level and 203 hours of computational time as the CPU-based method, the required computational resources would be approximately 323 PFLOPS (since A100 has a single-precision performance of 19.5 TFLOPS), which amounts to only $21.5\%$ of the supercomputer's overall resources. Furthermore, if mixed precision is allowed, which does not introduce significant errors, the computation resources required could be further reduced to 121 PFLOPS, i.e., only $8\%$ of the supercomputer.

We employed our algorithm to estimate the computational cost and time required to compute all 3 million sampled bitstrings from the 53-qubit 20-cycle Sycamore circuits. This task was accomplished using a CPU-based method~\cite{liu2022Validating} that utilized the entire supercomputer consisting of 41,932,800 cores with a single-precision peak performance of 1.5 EFLOPS, and the final validation task was completed in 203 hours.  A summary of our verification details is presented in Table~\ref{tab:n53m20_val_time}. Based on our computing complexity and efficiency, if we use the same single-precision peak performance as the CPU-based method, the required computing time would be 43.7 hours, only 21.5\% of the CPU-based approach. Moreover, if mixed precision is allowed, which does not introduce significant errors, the required computing time is further reduced to 16.4 hours, which is only 8\% of the supercomputer's time.

\begin{table}
\centering
  \caption{Verification details about 53-qubit 20-cycle Sycamore circuits.}
  \label{tab:n53m20_val_time}
  \begin{tabular}{l|lll}
    \hline
    % Computing complexity & $1.14\times 10^{22}$ \\
    Overall $T_{cc}$ & $9.15\times 10^{22}$\\
    Number of sub-tasks & $2^{30}$ \\
    Largest intermediate tensor & 32 GB \\
    One sub-task time (FP32 before optimization) & 17.39 sec. \\
    One sub-task time (FP32 after optimization) & 11.28 sec. \\
    One sub-task time (mixed after optimization) & 4.22 sec. \\
    \hline
\end{tabular}
\end{table}

We compared our simulation approach with other GPU-based simulation techniques for the task of generating bitstring samples with bounded fidelity in the RCS experiments on the 53-qubit 20-cycle Sycamore circuits. In the tensor network simulation of RCS experiments, all sub-tasks are orthogonal to each other~\cite{markov2018Quantum, pan2022Solving}. Thus, summing the fraction equal to the LXEB value of overall sub-tasks results in the bitstring samples with the same LXEB fidelity. Therefore, we can estimate the resources required for the simulation task when compared to others by multiplying a specific LXEB fidelity by our computational cost.

\begin{table}
\centering
  \caption{Cost of bounded LXEB fidelity ($0.224\%$) simulation task of 53-qubit 20-cycle Sycamore RCS experiments.}
  \label{tab:n53m20_sim_time}
  \begin{tabular}{lllll}
    \hline
    Methods &  Pan et al.\cite{pan2022Solving} & Kalachev et al.\cite{kalachev2021Classical}  & Our & Our\\
    \hline
     Overall $T_{cc}$ & $6.19\times 10^{18}$ & $2.2\times 10^{19}$ & $1.24\times 10^{19}$ & $2.56\times 10^{19}$\\
     \hline
    Device & V100-32GB & V100-16GB & V100-16G & A100-80GB\\
    % \hline
    %  performance (TFLOPS) & 3.08 & 4.86 & 20.20 \\
    \hline
     GPU hours & 4,604 & 10,800 & 2966 & 2,819\\
     \hline
     Bitstring num & $2^{20}$ & 2 million & 2 million & 3 million\\
    \hline
\end{tabular}
\end{table}

Table~\ref{tab:n53m20_sim_time} provides a comprehensive comparison of our simulation cost with the results reported in \cite{pan2022Solving, kalachev2021Classical}. Notably, our simulations generate a higher number of bitstrings, resulting in a computational cost that is up to 4.13x higher than the other two methods. However, our method's simulation cost is significantly lower, comprising only $61\%$ of the cost reported in \cite{pan2022Solving} and $26\%$ of the cost reported in \cite{kalachev2021Classical}. With the same overall computing complexity ($T_{cc}$), the simulation could be accelerated by 6.78 and 4.48 times, respectively. For fair comparison, the estimated $T_{cc}$ and GPU hours with the same bitstring number and device in~\cite{kalachev2021Classical} is also listed, the results indicate that the simulation time is only $27.46\%$ with the similar amount of computational complexity.

\section{Conclusions and Outlook}
\label{section:conclusions}
%%	Conclusion  
In this article, we propose several optimizations for tensor network simulation methods to accelerate sampling-based quantum experiments on large-scale, highly entangled quantum circuits using modern GPUs while maintaining simulation fidelity. Our optimizations include optimizing computing kernels and data precision. We convert sparse einsum operations into GEMM operations, reducing the time for the most time-consuming part of the overall simulation. To meet the high precision requirements of the LXEB verification task, we use extended precision to mitigate inaccuracies caused by using computing precision in customized processing units designed to accelerate tensor operations. In addition, we use mixed precision to accelerate large tensor operations in the simulation, and validate that this usage only has negligible effects on the final results. Combining these optimizations, we successfully verified RCS experiments on 53-qubit 18-cycle Sycamore quantum circuits, sampling over 1 million bitstrings with a 3.93x performance improvement compared to the non-optimized version. We also estimated the computational costs of the bounded LXEB simulation task and the verification task of 53-qubit 20-cycle Sycamore RCS experiments using our algorithms, and found that our final results were superior to other CPU and GPU-based methods.

%%	Outlook 
% For future work, we suggest exploring two avenues for optimizing tensor network simulations. The first direction is to optimize the permutation operations involved in tensor contraction. These operations can take a significant amount of running time during the simulation. While they are essential to the nature of tensor network contraction, it may be possible to reorder the modes of the initial and intermediate tensors to improve computing efficiency. The second direction involves re-scaling the quantum circuit data. Due to the unitary nature of quantum circuits and the exponential size of the Hilbert space, the floating-point numbers inside tensors tend to become increasingly smaller during contraction. To fully utilize the representation range of floating-point numbers and avoid errors due to floating-point truncation, a possible solution is to re-scale the tensors at the beginning or during the tensor network contraction process.

For future work, we suggest exploring two avenues to optimize tensor network simulations. The first is to optimize the index arrangement of the tensors involved from the perspective of improving the overall efficiency of tensor network contraction. This can bring a more regular memory access pattern without affecting the overall computational complexity, and can better make use of the full power of the Tensor Core. The second direction involves re-scaling the quantum circuit data. Due to the unitary nature of quantum circuits and the exponential size of the Hilbert space, the floating-point numbers inside tensors tend to become increasingly smaller during contraction. To fully utilize the throughput of low floating-point precision, a possible solution is to re-scale the tensors at the beginning or during the tensor network contraction process.
\begin{acks}
    We thank Daochen Shi, Rita Zhang, and Guofeng Zhou for helpful discussions.
    A python implementation that can reproduce the optimizations achieved in this paper is available at~\cite{implement}. The contraction order is generated by package ArTensor~\cite{artensor}. F. P. acknowledges the support from National Research Foundation of Singapore under its grant NRF2020-NRF-ISF004-3528. 
    P.Z. acknowledges the Project 11747601 and 11975294 of NSFC.
\end{acks}

\bibliographystyle{ACM-Reference-Format}
\bibliography{Main-ref}

% \iffalse
\appendix

\section{Research Methods}

\subsection{Unifying tensor network simulation methods into sparse-state simulation}

\begin{figure}[h]
    \centering
    \includegraphics[width=\linewidth]{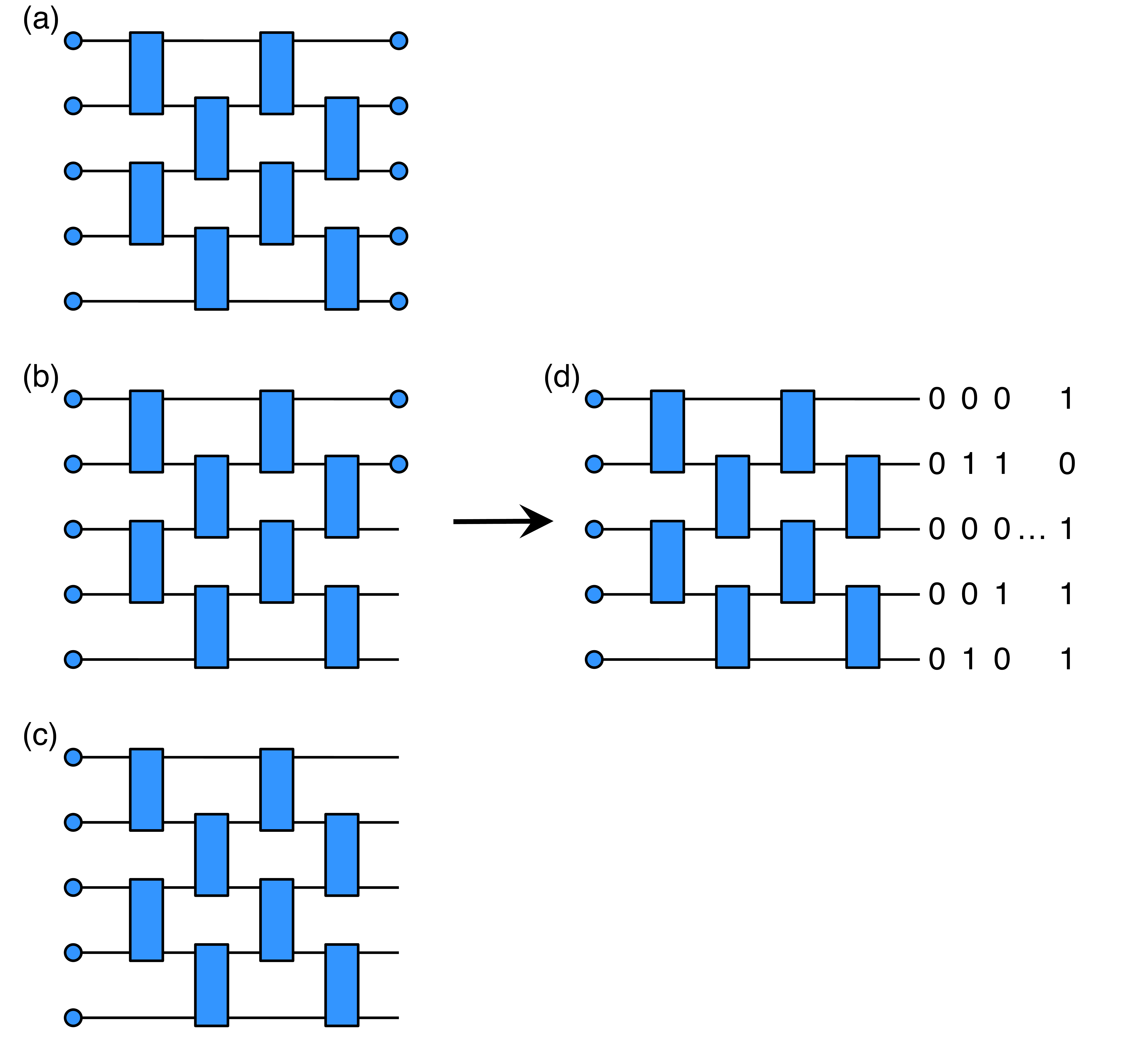}
    \caption{Unification of three traditional tensor network simulation methods into the sparse-state simulation.}
    \label{fig:unifying}
    \Description{}
\end{figure}

In the tensor network simulation of quantum circuits, there are typically three categories of simulation: single amplitude, full state, and subspace simulation.
For single amplitude cases, only the amplitude of a single bitstring is of interest. This bitstring is re-expressed as a product state, and then added to the end of the quantum circuit to construct a closed tensor network. The resulting tensor network is then repeatedly prepared to obtain amplitudes for different bitstrings using the single amplitude simulation method~\cite{markov2008Simulating}.
In contrast, full state simulation aims to obtain all amplitudes associated with the entire Hilbert space without setting any constraint in the final state. In this case, all bonds in the final state will be open and the contraction of such tensor network will output a $2^N$ vector which represents all amplitudes in the Hilbert space. However, the memory requirement of this approach is exponential in the number of qubits, rendering it feasible only for quantum circuits with a small number of qubits.
Subspace sampling lies between these two approaches. In subspace simulation, the final state is partially closed by applying product tensors to qubits that are not of interest~\cite{huang2021Efficient, pan2022Simulation}.
This allows the amplitudes of the opened Hilbert subspace to be directly obtained by contraction.
Subspace sampling is mainly used to obtain samples from quantum circuits with a large number of qubits, where the final state cannot be stored by the available computational resources. In such cases, the state of the closed qubits is randomly chosen and a bitstring is sampled from the opened subspace for further experimentation.

The main difference between the three tensor network simulation methods in quantum circuits is the level of restriction placed on the final state.
In single amplitude simulation, the restriction is maximal, as the final state is only allowed to represent a single bitstring.
In full state simulation, there are no restrictions on the final state and the entire information of the Hilbert space can be computed.
The amount of restriction placed on the final state depends on the size of the circuit and the computational resources.
The resulting tensor networks of these three methods are either closed tensor networks or tensor networks with few open bonds. The contraction of such tensor networks is typically ordinary and can be translated into a sequence of einsum equations.

% In this article, the main purpose is to calculate the amplitudes of multiple bitstrings sampled from quantum circuits' encoded final states. Due to the large number of qubits and the uncorrelated nature of the sampled bitstrings, full-state and subspace simulation are not viable options. However, if we use single amplitude simulation and repeat it for the number of bitstrings times, it is not efficient, and the cost of such repetition is huge for large circuit simulation.

In the main article, we used a different method called sparse-state simulation proposed in~\cite{kalachev2021Recursive,pan2022Solving,liu2022Validating} that is designed to calculate the amplitudes of multiple bitstrings in one tensor network contraction. Compared to the above three methods, sparse-state simulation adds a different kind of final state restriction, which only allows the final state to be the configuration in bitstring samples. This restriction acts as a boundary condition onto the final state and guides the open bond of qubits and when these bonds merge together.

% For example, if an open bond representing qubit $i$ is to be merged with another open bond representing qubit $j$, the initial open bond dimension of these two bonds will be 2. During the merge, they will refer to the sparse-state boundary condition to detect how many different configurations there are in the qubits $i$ and $j$ of the sparse state. If the number of unique configurations is 4, that means all possible states ($|00\rangle$, $|01\rangle$, $|10\rangle$, and $|11\rangle$) are contained in the sparse state; thus, the merge is ordinary, and the operation will be tensor product. 
% However, if the number of unique configurations is smaller than 4, this will lead to missing entries of the merged bond, meaning that if we do the contraction, we need to align the configurations of the merging bonds to output the exact unique configurations of unified Hilbert space.
% A pictorial illustration of such contraction can be found on Fig.

In Figure~\ref{fig:unifying}, we show that the sparse-state tensor network simulation framework unifies the above three methods. For single amplitude simulation, there will be only one state in the sparse-state boundary condition; thus, the number of unique configurations for each open bond merging will be 1. For full-state simulation, the final sparse-state boundary condition will be all states in the Hilbert space, which means that during each merging, the operation will always be a tensor product, and no index alignments are needed. And for subspace simulation, the circumstances will be a combination of the previous two cases. For qubits closed, the number of unique configurations will be 1, while for qubits opened, the number of unique configurations will be the number of states in such Hilbert subspace.

Since this method is so flexible and can encompass all previous cases, one may wonder why we called it the "sparse-state" simulation method. For the NISQ era, the number of qubits in quantum devices is large, which is beyond the full-state simulation capability of classical computers. At the same time, the number of sampled bitstrings from the quantum experiments will also be limited to a relatively small fraction of the entire Hilbert space because it is impossible to repeat the quantum experiments exponentially many times. Thus, the samples will definitely be a sparse state, and to calculate the probabilities of these samples, the boundary condition onto the final state will only be a constant number. Therefore, for the simulation of such typical quantum experiments, the overall method is dubbed sparse-state simulation.

\subsection{Balanced indices}
\begin{figure}[h]
    \centering
    \includegraphics[width=\linewidth]{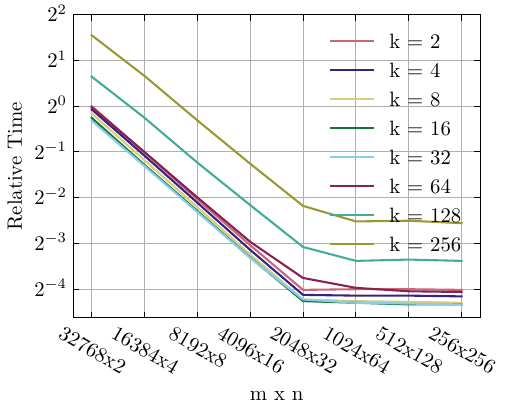}
    \caption{Comparison of actual execution time for batched GEMM with different shapes using TF32. The baseline shape for time is $1024\times32768\times2\times2$(batch size$\times$m$\times$n$\times$k), and the API being tested is \textit{cublasGemmBatchedEx()}.}
    \label{fig:chap4_bgemm}
    \Description{.}
\end{figure}

Finally, in the optimization process, we found that although some einsum operations can be converted into GEMM operation, they cannot fully tap into the computational efficiency of the GPU. Further analysis revealed that this is caused by an imbalance between the contracted indices and the remaining indices. When an einsum is converted into a GEMM(e.g., $C_{mn}=A_{mk}B_{kn}$), its contracted indices can be equivalent to $k$ in GEMM, and the remaining indices can be equivalent to $m$ and $n$, i.e., $m = {\prod \limits_{i=1}^{d_A} \text{len}(\alpha_{i})}/{\prod \limits_{i=1}^{d_S} \text{len}(\delta_{i})}$, $n = {\prod \limits_{i=1}^{d_B} \text{len}(\beta_{i})}/{\prod \limits_{i=1}^{d_S} \text{len}(\delta_{i})}$ and $k = \prod \limits_{i=1}^{d_S} \text{len}(\delta_{i})$ in Equation~\ref{eq:chap2_einsum}. When the dimensions $m$ or $n$ of the matrix are smaller than a certain value and there is a significant difference between them, the lower arithmetic intensity results in lower computational efficiency. As a result, the time for accessing matrices is much greater than the time for computing. When the dimension $k$ is small, global memory latency is more severe because pipelined parallelism is not well-suited in the $k$ direction. Therefore, in practical applications, it is necessary to try to keep $m$, $n$, and $k$ as similar or as large as possible. Figure~\ref{fig:chap4_bgemm} compares the actual relative execution time of different shapes of GEMM using Tensor Cores. The base shape for comparison is $32768\times2\times2$. It can be observed that when $m$ and $n$ are fixed, $k$ should be equal to or greater than 64 to fully utilize Tensor Cores. Similarly, when $k$ is fixed, $m$ or $n$ should be greater than or equal to 32 to fully utilize Tensor Cores. If $m$ or $n$ is less than 32, the closer the values are, the more efficient the Tensor Cores calculation will be. This shows that the matrix size needs to reach a certain threshold to fully utilize the computational power of Tensor Cores. To achieve this, we incorporated positive feedback into the path search algorithm. This resulted in a more balanced ratio, which significantly improves the efficiency of converted GEMM calculations. Additionally, it increases the computational efficiency of more balanced metrics for einsums that cannot be converted to GEMM.

% \fi


% ...


% ...

\end{document}